# Structure and dynamics of Oxide Melts and Glasses : a view from multinuclear and high temperature NMR

Dominique Massiot[1*], Franck Fayon[1], Valérie Montouillout[1], Nadia Pellerin[1], Julien Hiet[1], Claire Roiland[1], Pierre Florian[1], Jean-Pierre Coutures[2], Laurent Cormier[3] and Daniel R. Neuville[4]

1 - Centre de Recherches sur les Matériaux à Haute Température, CRMHT-CNRS, UPR4212, 1D avenue de la Recherche Scientifique, 45071 Orléans cedex 2, France
2 – PROMES-CNRS, Perpignan, France
3 – IMPMC, CNRS UMR 7590, Université Paris 6, Université Paris 7, 140 rue de Lourmel, 75015, Paris, France
4 – Physique des Minéraux et Magmas, CNRS-UMR 7154 IPGP, Paris, France

**Abstract**

Solid State Nuclear Magnetic Resonance (NMR) experiments allow characterizing the local structure and dynamics of oxide glasses and melts. Thanks to the development of new experiments, it now becomes possible to evidence not only the details of the coordination state of the network formers of glasses but also to characterize the nature of polyatomic molecular motifs extending over several chemical bonds. We present results involving $^{31}$P homonuclear experiments that allow description of groups of up to three phosphate units and $^{27}$Al/$^{17}$O heteronuclear that allows evidencing $\mu_3$ oxygen bridges in aluminate glasses and rediscussion of the structure of high temperature melts.

**Introduction**

Oxide glasses are known and used for thousands of years and tuning of properties like colour, durability, viscosity of the molten state were mostly known and dominated by glass makers. Despite this millenary knowledge, the range of glass forming system of interest is still expanding and many non-elucidated points remain in the understanding of the glass and melts structure and properties. The aim of this contribution is to underline, from the experimental point of view provided by Nuclear Magnetic Resonance, the relations existing between the structure and dynamics of the high temperature molten oxide systems and the short and medium range order of their related glasses.

---

[*] To whom correspondance should be adressed massiot@cnrs-orleans.fr

The strength of Nuclear Magnetic Resonance for describing structure and dynamics of amorphous or disorganised system like oxide glasses or melts, comes firstly from its ability to selectively observe the environment of the different constitutive atoms (providing that they bear a nuclear spin) and secondly from its sensitivity to small variation in the first and second coordination sphere of the observed nucleus. This often provides spectral separation of the different types of environment[1]. The information derived from NMR experiments are then complementary to those obtained by other means : optical spectroscopies, IR or Raman, X-Ray absorption, X-Ray or neutrons elastic or inelastic scattering etc… It is important to remark that NMR has a much slower characteristic time (ranging from Hz to MHz) than most of the above mentioned methods, leading to fundamental differences in the signatures of the viscous high temperature molten states.

**One dimensional NMR experiments**

In liquid state in general, and in the high temperature molten state in the case of oxide glass forming systems, the mobility is such that only the isotropic traces of the anisotropic interactions express in their NMR spectra. Fluctuation of these interactions leads to relaxation mechanisms that can allow discussion of the characteristic times of rearrangement and overall mobility of the system. In solid state materials and in glasses the anisotropy of the different interaction fully express in the static NMR spectra giving broad and often featureless line shapes accounting for all the different orientations of the individual structural motifs of the glass. Although these broad spectra contain many different information on the conformation of the structural motifs (Chemical Shift Anisotropy - CSA), spatial proximity between spins (homo- and hetero-nuclear Dipolar interactions), chemical bonds (indirect J coupling), electric field gradient at the nucleus position (Quadrupolar interaction for I>1/2 nuclei), these information are often hardly evidenced. Magic Angle Spinning is this unique tool that solid state NMR has at hand to average out all (or most) of the anisotropic part of the interactions only leaving their traces mimicking (or giving a coarse approach of) the Brownian reorientation of the liquid phase. Under rapid Magic Angle Spinning, Chemical Shift is averaged to its isotropic value and distribution directly given by the line position and width in the case of a dipolar (I=1/2) spin, while the traceless Dipolar interaction is averaged out, and the scalar (or isotropic) part of J-coupling is usually small enough to be completely masked in a 1D spectrum, even in crystalline phases.

Phosphates, silicates, alumino-silicates or aluminates oxide glasses structures are mostly based on tetrahedral species whose polymerization is characterized by their number of bridging oxygens ($Q^n$ : Q=P,Si,Al and n the number of bridging oxygens). Figure 1 presents the $^{31}$P MAS NMR 1D spectra of a (60% PbO-40% $P_2O_5$) glass. It shows two broad but resolved resonances in a 1/1 ratio that can unambiguously ascribed to end-chain groups ($Q^1$ 750Hz 6.2 ppm width) and middle-chain groups ($Q^2$ 1100Hz 9 ppm width) environments. Both these lines are considerably broader than that of the corresponding crystalline sample ($Pb_3P_4O_{13}$ linewidth < 1 ppm) due to the disorder in the glass structure and the loss of long range order. In the case of *simple* binary glasses of phosphates or silicate the broad lines corresponding to the various $Q^n$ tetrahedral sites are often resolved enough to allow quantification of their respective abundance and evaluation of the disproportionation equilibrium constants ($K_n$ : $2Q^n<->Q^{n-1}+Q^{n+1}$)[2]. Figure 2 reports these quantitative results for PbO-$SiO_2$[3] and PbO-$P_2O_5$[4] binary systems. In lead-phosphate glasses the $K_n$ values remain very small which correspond to a binary distribution and indicates that only two types of $Q^n$ environments can co-exist at a given composition, while in lead-silicate glasses the equilibrium constant are much higher, close to that of a randomly constructed network with a competition between lead based and silicon based covalent networks. $^{207}$Pb NMR and $L_{III}$-EXAFS experiments confirmed this interpretation by showing that the coordination numbers of Pb in silicate is of 3 to 4 oxygen with short covalent bonds and a very asymmetric environment (pyramid with lead at the top) while it is of more than 6 in phosphate glasses with a more symmetric environment, behaving more as a network modifier[3-5].

**Polyatomic *molecular* motifs**

Although these information already give important details on the structure of these phosphates or silicate binary glasses, it would be of great interest to obtain a larger scale image of the polyatomic *molecular* motifs constituting these glasses and especially to evaluate the length of phosphate chains possibly present in the glass, that makes the difference between the long range ordered crystalline phase and the amorphous phase. That type of information can be obtained by implementing multidimensional NMR experiments that allow to evidence Dipolar[4] or J-coupling[6,7,8] interaction and further use them to build correlation experiments separating the different contributions of well defined *molecular* motifs. Figure 1 gives a general picture of the possibilities offered by the J-coupling mediated experiments that allow to directly evidence the P-O-P bonds bridging phosphate units through $J^2_{P-O-P}$ interaction. Let us consider the example of the 60% PbO-40% $P_2O_5$ glass already introduced above. Its 1D spectrum (fig 1a) shows partly resolved $Q^1$ and $Q^2$ lines with strong broadening (750 and 1100Hz) signing the *to be understood* glass disorder. Because the $Q^1$ and $Q^2$ line width is essentially inhomogeneous, due to distribution of frequencies for each individual motif, this broadening can be refocused in an echo which is modulated by the small (and unresolved) isotropic $J^2_{P-O-P}$ coupling[7]. Figure 1b shows the J-resolved spectrum of the glass that reveals the J coupling patterns consisting in a doublet for $Q^1$, and to a triplet for $Q^2$, thus justifying the spectral attribution previously made based on the $^{31}$P isotropic chemical shift. It is also of importance to notice that this experiment clearly shows that the isotropic J-coupling does vary across the 1D lines, typically increasing with decreasing chemical shift. The new type of information provided by this experiment is likely to bear important information on the covalent bond hybridisation state and geometry. Because this isotropic J-coupling can be measured, it can also be used to reveal - or to spectrally edit - different polyatomic *molecular* units in the glass. Figure 1c and 1d respectively show the two-dimensional correlation spectra that enable the identification of through-bond connectivity between two linked $PO_4$ tetrahedra (fig.1c) [6] and between three linked $PO_4$ tetrahedra (fig.1d) [8]. These experiments, fully described in the referenced papers allow spectral separation of dimers, end-chain groups, and middle-chain groups when selecting Q-Q pairs (fig.1c) and trimers, end-chain triplets or centre-chain triplets when selecting Q-Q-Q triplets (fig.1d). From these experiments it becomes possible to identify the different structural motifs constituting these glasses in terms of *molecular* building blocks extending over 6 chemical bonds (O-P-O-P-O-P-O) over lengths up to nearly 10Å if we consider a linear chain. Other experiments of the same type now allow to describe hetero-nuclear structural motifs of different types involving Al-O[9], Al-O-Si, P-O-Si[10] or opening the possibilities of more detailed description of glasses or disordered solids at large length scale.

**High Temperature NMR experiments**

Even if most of the resolution is lost when going to static NMR spectra in the general case, the very different chemical shift anisotropy of $Q^3$ and $Q^4$ silicon environment can be source of enough resolution for evidencing dynamic process occurring close to or above glass transition temperature as shown by Stebbins and Farnan in the case of a binary $K_2O-4SiO_2$ composition[11]. They showed that while the two $Q^3$ and $Q^4$ contributions can be resolved from their different chemical shift anisotropy or from their isotropic chemical shift, below glass temperature, they begin to exchange just above glass transition with characteristic times of the order of seconds[12] and finally end up into merging in a unique line in the high temperature molten state. This experiment underlines two important points. First, although silicate glasses can be regarded as $SiO_2$ based polymer, the melting of silicate glasses implies rapid reconfiguration of the structural motifs through a mechanism that was proposed to involve a higher $SiO_5$ coordination state of silicon with oxygen, second that the characteristic time scales of NMR spectroscopy allow to explore a large range of time scales involved in this mechanism. We can remark that this has been recently extended to below $T_g$ structural reorganisation of $BO_3$ and $BO_4$ configurations in borosilicate glasses[13]. The existence of higher (and previously unexpected) $SiO_5$ coordination state of silicon was proved experimentally by acquiring high quality $^{29}Si$ NMR spectra[14] with clear effects of quench-rates and pressure stabilizing these high coordination silicon environments.

The high temperature NMR setup developed in our laboratory, combining $CO_2$ laser heating and aerodynamic levitation allows acquisition of $^{27}Al$ resolved NMR spectra in molten oxide at high temperature with a good sensitivity[15,16]. Figure 3a shows the experimental setting and an example of a $^{27}Al$ spectrum acquired in one scan for a liquid molten sample $CaAl_2O_4$ at ~2000°C[17]. The sensitivity of this experiment is such that one can follow in a time-resolved manner the evolution of the $^{27}Al$ signal when cooling the sample from high temperature, until disappearance of the signal when the liquid becomes too viscous. As in the case of the high temperature molten silicate discussed above, we only have a single sharp line giving the average chemical shift signature of the rapidly exchanging chemical species. This later point is confirmed by independent $T_1$ (spin-lattice) and $T_2$ (spin-spin) relaxation measurements giving similar values and reliably measured in the 1D spectrum from the linewidth. This relaxation time can be modelled using a simple model of quadrupolar relaxation which requires the knowledge of the instantaneous quadrupolar coupling that can be estimated from the $^{27}Al$ MAS NMR spectrum of the corresponding glass at room temperature. The obtained correlation times, corresponding to the characteristic time of the rearrangement of aluminium bearing structural units, can be directly compared to characteristic times of the macroscopic viscosity with a convincing correspondence in the case of aluminates melts[18] (Figure 3b&c).

**Structure and dynamics of alumino-silicates**

In alumino-silicate glasses of more complex composition, aluminium is able to substitute silicon in tetrahedral network forming positions, providing charge compensation by a neighbouring cation. In such case, the NMR signature of $^{29}$Si spectra is much more complex and difficult to interpret[19]. Because $^{29}$Si $Q^n$ species isotropic chemical shifts depend upon Al substitution in neighbouring tetrahedra, $^{29}$Si silicon spectra are usually broad Gaussian lines covering the full range of possible environment. Similarly $^{27}$Al aluminium spectra are broadened by the combination of a distribution of chemical shifts and a distribution of second order quadrupolar interaction[21] and give only average pictures of the structure with possible resolution of different coordination states but no resolution of the different Al based $Q^n$ species except in the case of binary $CaO-Al_2O_3$ glasses in which NMR and XANES both show proofs of the depolymerization of the $AlO_4$ based network[20]. In alumino-silicate glasses, aluminium species with higher coordination were evidenced[22] and quantified[21] using a detailed modelling of the $^{27}$Al MAS and MQMAS NMR spectra obtained at high principal fields. One can also remark that no $SiO_5$ environments have ever been evidenced in aluminosilicate compositions. Going further and examining the whole $SiO_2-Al_2O_3-CaO$ phase diagram[23], we showed that these $AlO_5$ environments are not confined to the charge compensation line or to the hyper-aluminous region of the ternary diagram, where there exist a deficit of charge compensators, but that $AlO_5$ species are present, at a level of ~5%, for any alumino-silicate composition of this ternary diagram, including those presenting the smallest fraction of alumina but excluding the Calcium Aluminates of the $CaO-Al_2O_3$ join which nearly exclusively show aluminium in $AlO_4$ coordination state. For C3A composition XANES unambiguously shows that Al occupy $Q^2$ environments both in crystal, and glass[20,23].

These finding that there exist no or very few $AlO_5$ in compositions close to $CaO-Al_2O_3$ composition is somehow in contradiction with our previous interpretation of chemical shift temperature dependence with a negative slope [17]. At that time we proposed to consider that there could exist significant amounts of five fold coordinated aluminium in the high temperature molten state, based on the thermal dependence of chemical shift and on state of the art MD computations. A more detailed study shows that, across the $CaO-Al_2O_3$ join, the slope of the temperature dependence of the average chemical shift in the high temperature molten state drastically changes from a positive value for $Al_2O_3$ to very negative (~-4 to -5ppm) for composition around $CaAl_2O_4$. Indeed we can even remark that all compositions able to vitrify in aerodynamic levitation contactless conditions have a slope smaller than -2ppm/1000°C (Figure 4a). Stebbins and coworkers recently studied the $^{17}O$ NMR signature of similar composition[24]. They evidenced a significant amount of non bridging oxygen atoms and discussed the possibility of a seldom observed $\mu_3$ tricluster oxygen linking three tetrahedral Al sites that exists in the closely related $CA_2$ ($CaAl_4O_7$ - Grossite) crystalline phase. Thanks to the development of new methods of hetero-nuclear correlation between quadrupolar nuclei through J-coupling at high principal field (750MHz)[9], we could re-examine this question and show that a $\{^{17}O\}^{27}Al$ experiment carried out on a $CaAl_2O_4$ glass was able to clearly evidence the signature of ~5% $\mu_3$ tricluster oxygen linked to aluminium with chemical shift decreased by 5 ppm per linked tricluster (Figure 4b). It thus appears that molecular motif of type $\mu_3[AlO_3]_3$ can be quenched in the glass and do exist in the molten state while $AlO_5$ remains negligible, rising a new interpretation of the thermal dependence of the $^{27}Al$ isotropic chemical shift in the $CaO-Al_2O_3$ melts.

**Conclusion**

From the above discussed experimental results we can draw several important points about the relations between structure and properties of oxide glasses and their related molten states which appear to be closely related. It first clearly appears that in many cases, even if most of the structure of the glasses, and consequently of their related high temperature molten states are built around a network of $\mu_2$ connected tetrahedra (P, Si, Al…), there exist in many cases unexpected environments showing up as minor contributions in the glass structures (~5% or less) but significantly present and relevant to molecular motifs that can be identified. This is the case of $SiO_5$ species in binary alkali silicates[14], $AlO_5$ ($AlO_6$) [21-23] in aluminosilicates, violations to Al avoidance principle[25] or tricluster $\mu_3$ oxygens[9]. This implies that modelling of these complex materials in their solid or molten state will often be difficult using limited box sizes. Just consider that 5% of Aluminum species in a glass containing 5% $Al_2O_3$ in a Calcium Silicate only represent 1 to 2 atoms over 1000 or that 5% $\mu_3$ oxygens in a $CaAl_2O_4$ composition represent less than 3 occurrences in a box of 100 atoms. Furthermore Charpentier and coworkers[26] recently showed that a proper rendering of NMR parameters from all electrons ab-initio computations in glasses requires a combination of classical and ab-initio MD simulation. Going further we also emphasize that an important part of what we qualify with the general term of *disorder* can be described in terms of distribution of poly-atomic *molecular* motifs extending over a much larger length scale than the usual concept of coordination.

**Acknowledgements**

We acknowledge financial support from CNRS UPR4212, FR2950, Région Centre, MIIAT-BP and ANR contract RMN-HRHC.

**Figure Captions :**

*Figure 1*   Summary of NMR experiments on a 60%PbO-40%P$_2$O$_5$ glass evidencing poly-atomic molecular motifs with : (a) 1D spectrum, (b) the J resolved spectrum showing doublet for Q$^1$ and triplet for Q$^2$ [7], (c) the INADEQUATE experiment evidencing pairs of phosphates (Q-Q)[6], and (d) the 3Quantum spectrum evidencing triplets of phosphates (Q-Q-Q)[8].

*Figure 2*   Quantitative interpretation of $^{29}$Si and $^{31}$P 1D spectra allowing the measurement of disproportionation constants for (a) lead silicate[3] and (b) lead phosphate glasses[4].

*Figure 3*   (a) High Temperature aerodynamic levitation NMR setup and a characteristic one shot spectrum, (b) temperature dependence of the chemical shift and (c) viscosity and NMR correlation times[adapted from ref.17]

*Figure 4*   (a) Slope of the thermal dependence of average chemical shift in high temperature versus composition for the CaO-Al$_2$O$_3$ join. (b) {$^{17}$O}$^{27}$Al HMQC experiment of CaO-Al$_2$O$_3$ glass at 750 MHz showing clear signature of µ$_3$ tricluster oxygens[adapted from ref.9].

# Figure 1

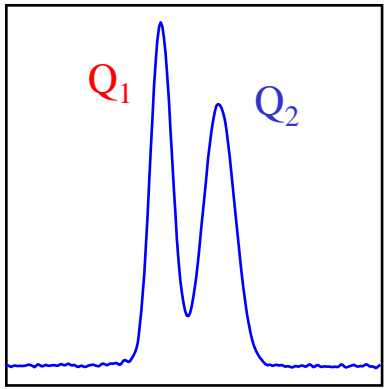

(a)

**$^{31}$P MAS NMR
In PbO-P$_2$O$_5$ glasses

Characterization
up to 4 chemical bonds
P-O-P-O-P**

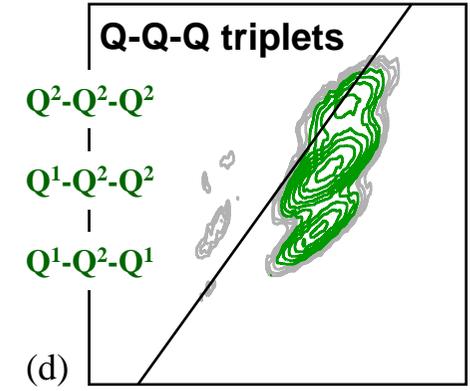

Q-Q-Q triplets

$Q^2$-$Q^2$-$Q^2$
$Q^1$-$Q^2$-$Q^2$
$Q^1$-$Q^2$-$Q^1$

(d)

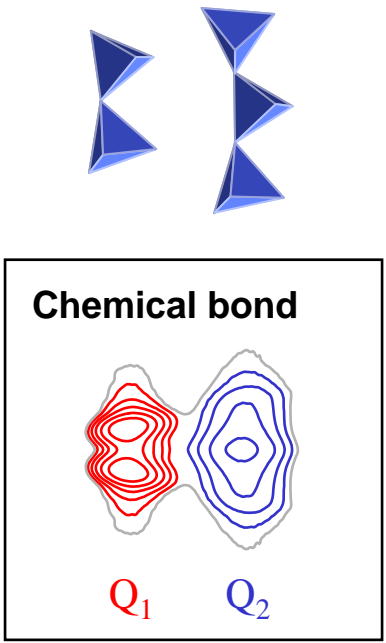

**Chemical bond**

$Q_1$  $Q_2$

(b)

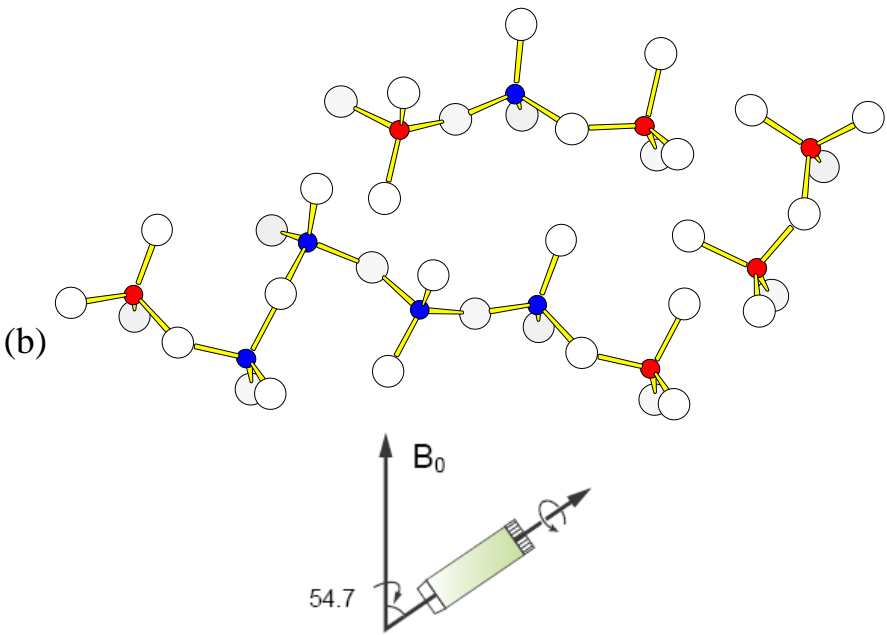

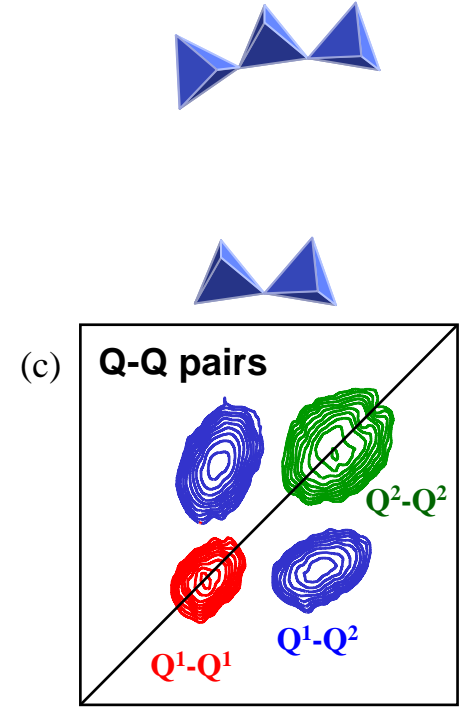

(c) Q-Q pairs

$Q^2$-$Q^2$
$Q^1$-$Q^2$
$Q^1$-$Q^1$

# Figure 2

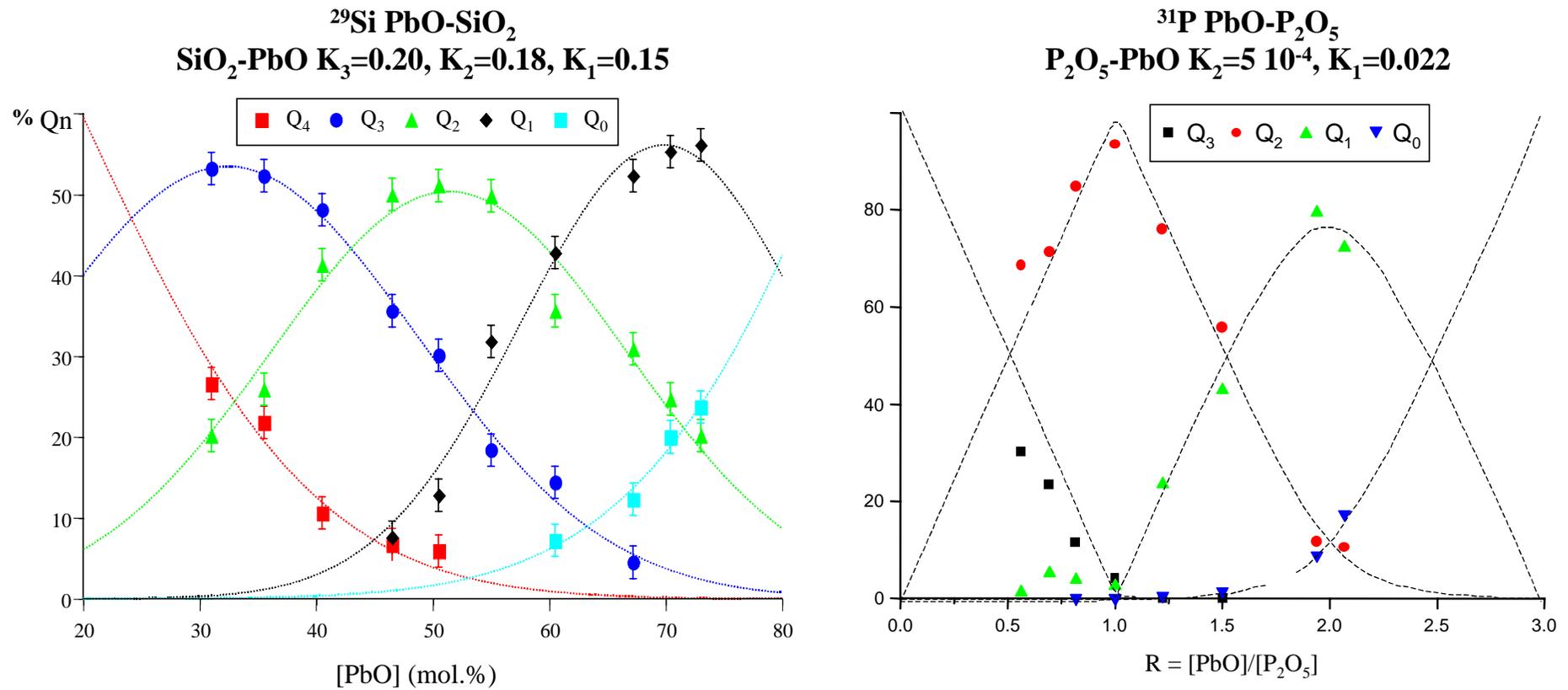

Disproportionation equilibrium
$2Q_n \leftrightarrow Q_{n-1} + Q_{n+1}$

# Figure 3

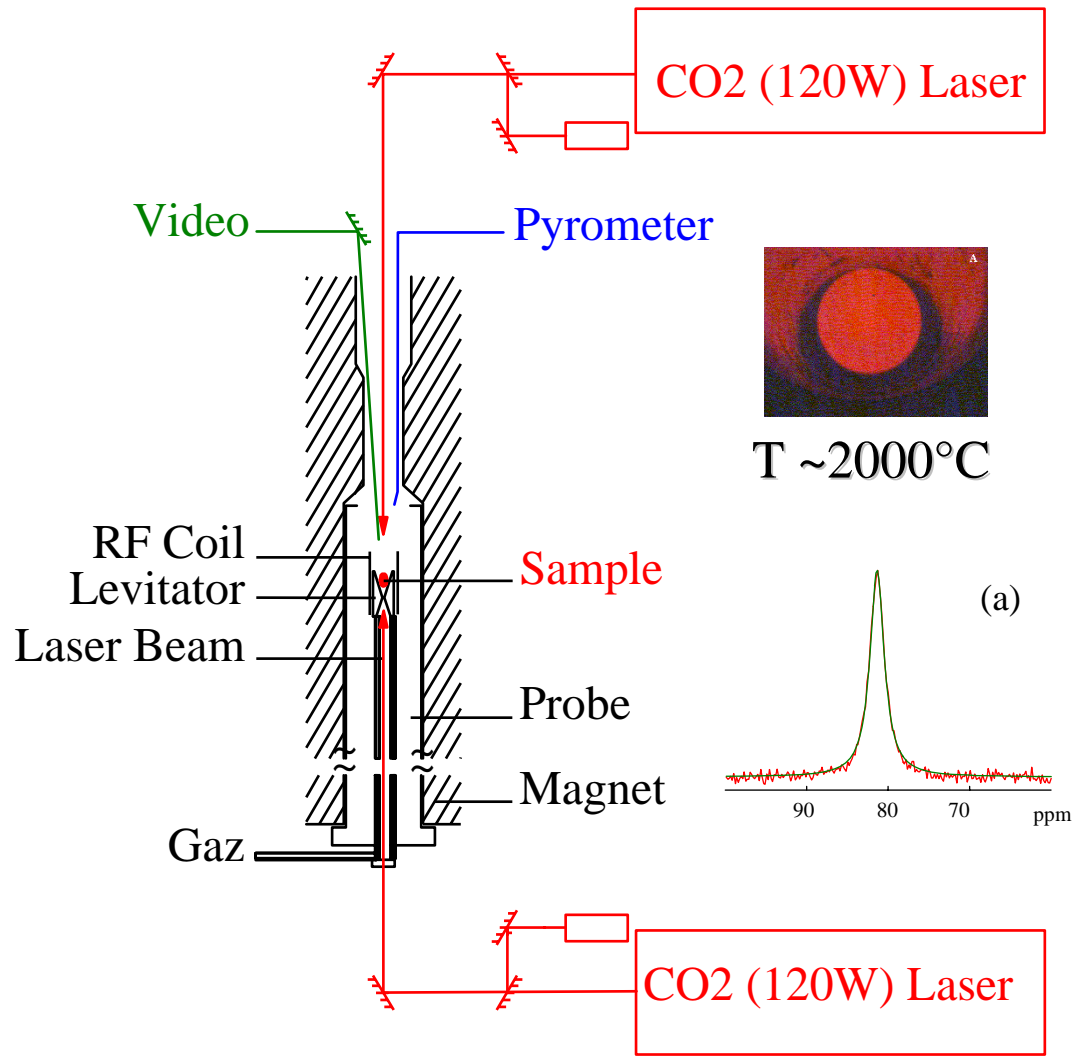
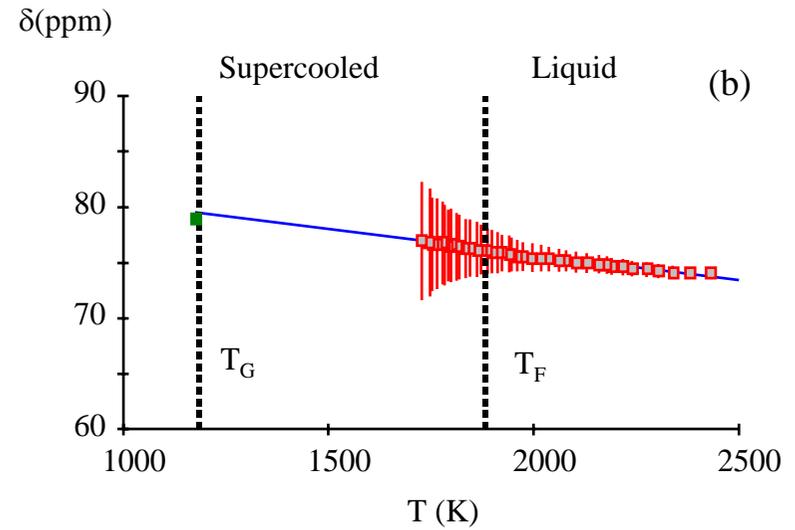
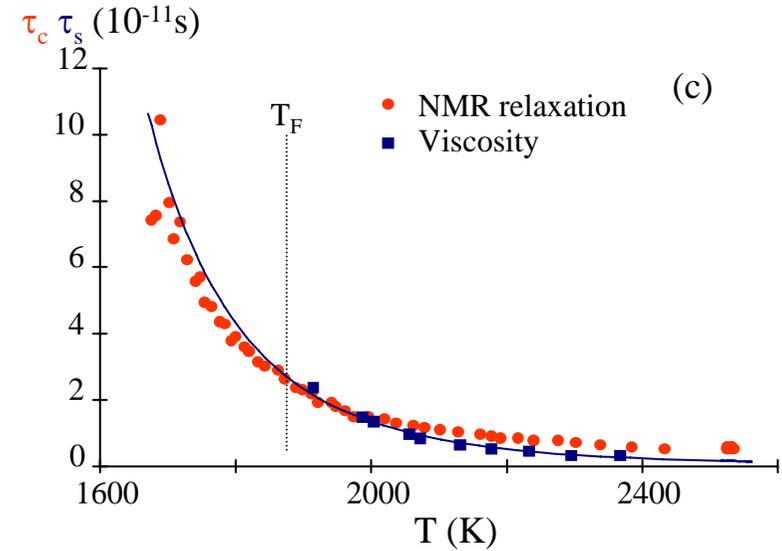

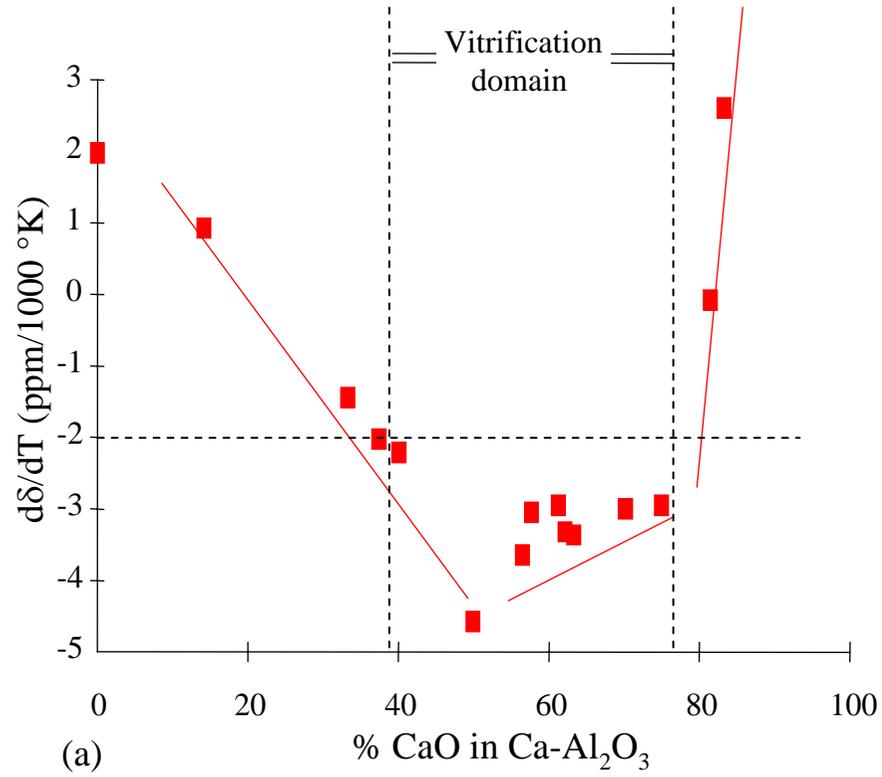
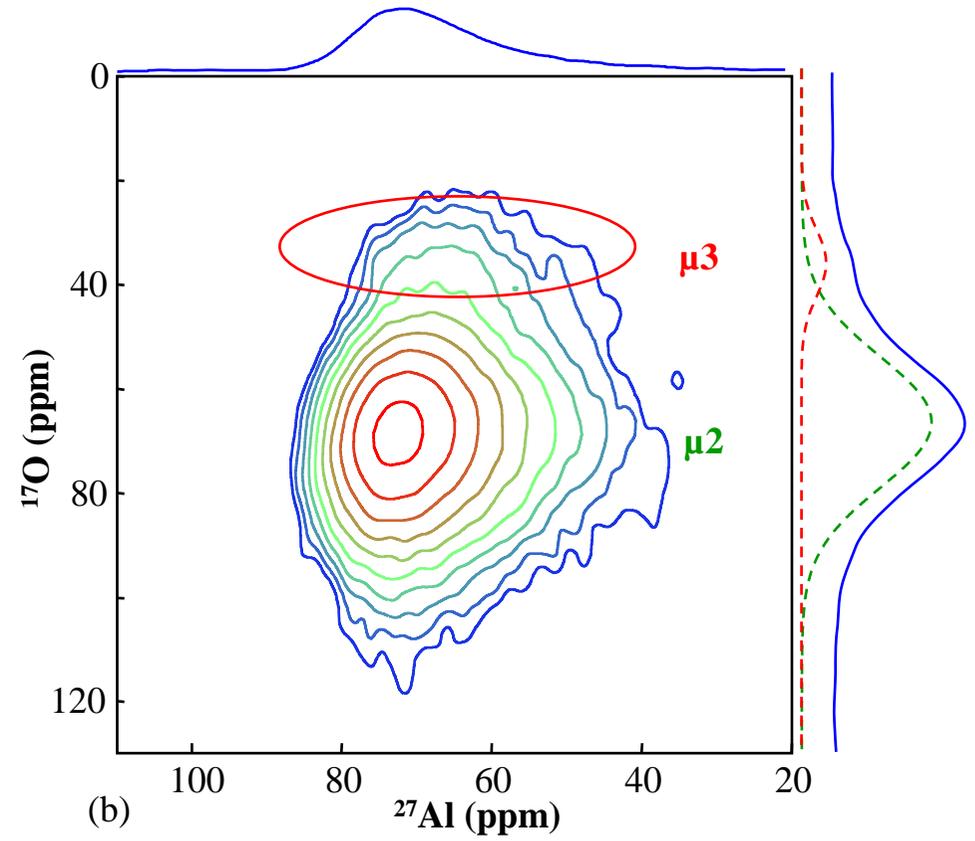

(a)  (b)